\documentclass[aps,pr,reprint,groupedaddress]{revtex4-1}
\usepackage[final]{graphicx}
\usepackage{dcolumn}
\usepackage{graphicx}
\usepackage{amssymb}
\usepackage{epstopdf}
\usepackage{amsmath}
\usepackage[footnotesize]{caption}

\graphicspath{%
    {converted_graphics/}
    {/}
    {gas4p2K/}
    {alphas/}
}
\usepackage[compact]{titlesec}

\begin{document}
\ \ \ \ \ \ \ \ \ \ \ \  \ \ \ \ \ \ \ \ \ \ \ \ \ \ \ \ \ \ \ \ \ \ \ \ \ \ \ \ \ \ \ \ \  \ \ \ \ \ \ 
\title{Charge Distribution about an Ionizing Electron Track in Liquid Helium}

\author{G. M. Seidel$^1$}
\email{george\_seidel@brown.edu}
\author{T. M. Ito$^2$}
\email{ito@lanl.gov}
\author{A. Ghosh$^{1}$}
\email{present address: Department of Electrical Communication, Indian Institute of Science, Bangalore, India}
\author{B. Sethumadhavan$^{1}$}
\email{present address: Intel Corporation, Portland, Oregon}
\affiliation{$^1$Department of Physics, Brown University, Providence, Rhode Island, 02912, USA\\
$^2$Los Alamos National Laboratory, Los Alamos, New Mexico 87545, USA}

\date{\today}

\begin{abstract}
The dependence on an applied electric field of the ionization current produced by an energetic electron stopped in liquid helium can be used to determine the spatial distribution of secondary electrons with respect to their geminate partners. An analytic expression relating the current and distribution is derived. The distribution is found to be non-Gaussian with a long tail at larger distances. \\ 

\end{abstract}

\pacs{XXXXXXXXXXXX}

\maketitle

\section{INTRODUCTION}\vspace{.2cm}

The nature of the distribution of electrons and positive ions about the track of an ionizing particle as it passes through matter has been of interest for many years, beginning with Jaff\'{e}\cite{Jaffe} who first discussed the influence of an electric field on the recombination of these charges. He assumed the ions and electrons formed two 
inter-penetrating columns of charges, the recombination of which is determined by the interaction of the columns with one another and an applied electric field. Later Onsager\cite{Onsager} pointed out that in many physical situations recombination is more likely to involve the interaction of a single pair of charges that are closest to one another rather than with the extended sea of other charges. Both the columnar and geminate approaches to recombination of charges have been complemented, criticized, modified, and extended in various ways in hundreds of papers in order to understand better the effects of ionizing radiation in a variety of materials. The recent use of the liquefied noble gases as radiation detectors for exotic particles with improved energy resolution has generated renewed attention to these problems\cite{snowmass}.   \\

We have been considering the possible use of liquid helium as a detection medium for neutrinos and dark matter particles\cite{helium,helium2, helium3,Ito}. Hence, we have become interested in the scintillation and in the recombination of charges about the track of an ionizing particle in liquid helium. This paper is directed to a discussion of what can be determined about the charge distribution about an electron track in helium from a measurement of the current as a function of an electric field applied to the liquid in which ionizing particles are stopped.\\

 Liquid helium has a number of advantages, when compared to other substances, as a medium in which to study the charge distribution about the track of an ionizing particle. The atomic physics of an electron collision with a helium atom has been well studied experimentally and theoretically. Similarly, the dynamics of a collision of a proton and an alpha particle with a helium atom has been investigated extensively, see Ref.~\cite{Ito} for references. The properties of thermalized electrons and helium ions are well known in the liquid\cite{Borgh}. An electron forms a cavity, or ``bubble" of 1.9 nm radius, for the reason that in so doing the exchange energy with electrons on neighboring atoms is decreased. A positive He$^+$ ion, on the other hand, first forms a dimer He$_2^+$ that becomes  the core of a solid ``snowball", because of electrostriction, with radius less than that of the electron bubble. \\

A consideration of the interaction between an electron and a positive ion in the presence of an applied electric field has led us to a means of determining the charge distribution produced by an electron track in liquid helium. The field dependence of the ionization current (the probability of escape from recombination in the language of Onsager) can be used to measure the charge distribution under the conditions that are applicable in liquid helium, in particular, a system where the charges move with a velocity proportional to the force applied. This requirement that the mobility of the charges is a constant, independent of electric field, while valid for localized charges in liquid helium, does not apply to the motion of delocalized electrons in the heavier liquefied noble gases or the motion of electrons with extended wave functions in gases and liquids, more generally. Delocalized electrons are heated by the electric field and as a consequence the momentum transfer cross section increases and the mobility decreases with field\cite{Huxley}.  \\

Furthermore, this analysis does not take into account the effects of diffusion, which is central in the Onsager treatment of geminate recombination. The absence of any regard for diffusion is justified in considering the motion of charges in liquid helium because of the low temperature, but this simplification limits the applicability of the present analysis to other systems.  That diffusion can be neglected for the motion of charges in liquid helium is shown to be valid later in this paper using the Onsager theory.\\

This paper is organized as follows. Section II develops an analytic technique for determining the charge distribution of secondary electrons produced by a primary ionizing electron. Section III discusses the charge distribution inferred from the current produced by a $^{63}$Ni beta source in helium, and Section IV summarizes the results of this analysis.\\
\vspace{.2cm}

\section{CHARGE DISTRIBUTION}\vspace{.2cm}

\subsection{Property of electron tracks}\vspace{.2cm}

The energy distribution of secondary electrons produced by the ionization of helium as a consequence of being hit by an energetic electron, proton or alpha particle has been determined in a number of experiments. In general, the number of secondaries decreases as their energy increases. For an energetic alpha particle the secondaries can have energies extending up into the several hundred eV\cite{Rudd92}. However, when the primary ionizing particle is an electron with energy below 100~keV, the secondaries rarely have energies above the first excitation level of 19.8 eV. The only mechanism by which these secondary electrons can lose energy and thermalize is by elastic scattering from helium atoms, a very inefficient process because of the mass difference between the electron and atom. The kinetic energy of an electron must drop to 1~eV or below before it becomes localized by forming a bubble. The fractional energy loss of an electron in a collision with an atom is the order of 
\begin{equation}
\frac{dE}{E}\approx \frac{2m}{M} \ ,
\end{equation}
where $m$ and $M$ are the masses of the electron and helium atom, respectively. Hence the number of collisions required to reduce the energy from, say, 10 eV to 1 eV is 
\begin{equation}
N\approx \frac{M}{2m} \ln(10)\approx 10^4.
\end{equation}
The elastic scattering cross section varies somewhat with energy of the electron varying from $3\times 10^{-16}$ cm$^2$  at 20 eV to $6\times 10^{-16}$ cm$^2$ at 1 eV\cite{Brunger}. Since the number density of liquid helium is $2\times 10^{22}$ cm$^{-3}$, the mean free path is the order of $\lambda \sim 10^{-7}$ cm and an electron in undergoing a random walk as it thermalizes will be roughly within a sphere $\sqrt{N}\lambda \sim 10^{-5}$ cm from the positive ion. \\

The stopping power\cite{ESTAR} and W-value of 43~eV of helium are such that
for an electron with energy of 17~keV (the mean energy of electrons from the $^{63}$Ni beta emitter used in this experiment) the average separation of ionization events in the liquid is $2.2 \times 10^{-5}$~cm.
 Thus, an electron is more likely to remain closer to its ionic partner than to other ions. The recombination is considered to be geminate. Once the electron is localized in a bubble, the ion and electron move in their opposite's Coulomb field and whatever applied field exists. The positive ion snowball, because of its smaller size,  has a somewhat larger mobility than the electron and accounts for most of the relative motion. Under typical conditions of temperature and applied field (the field is not so large or the temperature not so low that the moving charges create vortex rings) the charges move in a viscous medium with a velocity proportional to the force on them.\\
 
\subsection{Condition for charge separation}\vspace{.2cm}

When moving in a viscous medium, a pair of isolated charges of opposite sign will recombine or instead be separated by an applied electric field, depending on their initial separation and orientation with respect to the field. The condition for separation, derived in the appendix, is
\begin{equation}
r_0^2 (1+\cos\theta_0)\geq \frac{2e}{4\pi\epsilon_0\ \mathcal E}\ ,
\end{equation}
where $r_0$ is the initial separation of the charges, and $\theta_0$ is the angle the vector connecting the charges makes with the applied field ${\mathcal E}$. The critical assumption in deriving this equation is that the mobility is independent of velocity (electric field), that is, the parameter $\lambda$ is the same in Eqs. (A3) and (A4). This is based on the fact that the bubble and snowball are large objects with hydrodynamic masses the order of 100 helium atoms, and their motion in the normal fluid can be considered to be that of rigid spheres in a viscous continuum. Experimentally, Keshishev {\it et al.}\cite{keshishev} found in normal liquid helium no dependence of the mobility of either the bubble or snowball on field up to $10^4$~V/cm. \\

The distribution of secondary electrons with respect to their respective geminate partner, $D(r)$,  once they have thermalized in undergoing random elastic scattering, is taken to be independent of orientation and only dependent on distance. Then the fraction of electrons that escape recombination as a function of field is given by 
 
$$f=\int_0^{\infty} \int_0^{\arccos(2e/{4\pi\epsilon_0 \mathcal E}r^2-1)}\int_0^{2\pi} D(r) r^2 dr \sin(\theta) d \theta d \phi\ $$
\begin{equation}\Big/ \ 4 \pi\int_0^{\infty} D(r) r^2 dr
\end{equation}
 or
 \begin{equation}
 f=\int_{(\frac{e}{4\pi\epsilon_0 \mathcal E})^{1/2}}^{\infty} D(r) r^2 dr (1-\frac{e}{4\pi\epsilon_0 {\mathcal E}r^2})\ \Big/\ \int_0^{\infty} D(r) r^2 dr .
\end{equation}
\vspace{.2cm}

\subsection{Determination of charge distribution from field dependence of current}\vspace{.2cm}

If a beta source of moderate activity is placed inside a cell to which a uniform field can be applied then the steady state current produced by ionization is
\begin{equation}
 i({\mathcal E}) = 4\pi \int_{(\frac{e}{4\pi\epsilon_0 \mathcal E})^{1/2}}^{\infty} D(r) r^2 dr (1-\frac{e}{4\pi\epsilon_0 {\mathcal E}r^2})\ .
\end{equation}
The current is expressed as a fraction of the value it would have if all the charges were separated by the field.\\

Since the argument within the integral is zero at the lower limit, the derivative of current with respect to field is
\begin{equation}
 \frac{di({\mathcal E})}{d{\mathcal E}} = 4\pi \int_{(\frac{e}{4\pi\epsilon_0 \mathcal E})^{1/2}}^{\infty} D(r) \frac{e}{4\pi\epsilon_0 {\mathcal E}^2} dr\ . 
\end{equation} 
 Then
\begin{equation}
i+{\mathcal E}\frac{d i}{d{\mathcal E}} = 4\pi \int_{(\frac{e}{4\pi\epsilon_0 \mathcal E})^{1/2}}^{\infty} D(r) r^2 dr\  ,
\end{equation}
and its derivative 
\begin{equation}
\frac{d}{{d\mathcal E}}(i+{\mathcal E}\frac{d i}{d{\mathcal E}}) = 4\pi  D\Big( (\frac{e}{4\pi\epsilon_0 \mathcal E})^{1/2}\Big)  \frac{1}{2}\Big(\frac{e}{(4\pi\epsilon_0}\Big)^{3/2}\frac{1}{{\mathcal E}^{5/2}} \ .
\end{equation}
Hence, the dependence of current on field provides a means to obtain the charge distribution, $D(r)$, where $r = (e/4\pi\epsilon_0 {\mathcal E})^{1/2}$.
\begin{equation}
 D(r) = \frac{4\pi^{1/2}\epsilon_0^{3/2}{\mathcal E}^{5/2}}{e^{3/2}}\frac{d}{{d\mathcal E}}(i+{\mathcal E}\frac{d i}{d{\mathcal E}})\ .
\label{eq.dis}
\end{equation}
The charge distribution can be determined from the first and second derivatives of the current with respect to field.\\

\section{EXPERIMENT}\vspace{.2cm}

\subsection{Apparatus}\vspace{.2cm}

A measurement of current with respect to field was made in three different cells. All cells were cylindrical with electrodes on the ends of the cylinder. For measurements in the liquid, one cell had a diameter of 2.5~cm and a height of 0.4~cm (See Ref.~\cite{Seth06} for more details), and another had dimensions of 6~cm diameter and 1~cm height. For measurements in the gas, the cell had a diameter of 6 cm and height of 3.8~cm. The source of electrons was a 1~mCi $^{63}$Ni~$\beta$ emitter having an end point of 66~keV. The source was placed on a metal substrate that was part of one of the electrodes. The range of a 66~keV electron is $5\times 10^{-2}$~cm in 2.5~K liquid and is 0.6~cm in 4~K gas. Therefore, for all measurements all primary electrons
ranged out well within the cell. The saturation current, measured in  gas, was 2900~pA. The accuracy to which the current could be
determined varied somewhat with the measurement, ranging from 0.1~pA at low currents to 5\% of the value at high currents.
Measurements of current were made with both polarities of the field between the electrodes. 
The results for the two polarities were consistent with each other when account was taken for the contribution from the primary electrons, which was about 6.2~pA, corresponding to a source activity of 1~mCi. The absence polarity dependence indicates that certain possible systematic effects were small (see discussion below). The ratio between the saturation current and the current due to the primary electrons is consistent with the known $W$~value of 43~eV. The data for liquid taken with two different cells were consistent with each other within the uncertainty of the measurements where the data overlap. This indicates that geometry dependent systematic effects, such as those due to leakage currents, are smaller than the statistical uncertainty.\\

\subsection{Liquid helium}\vspace{.2cm}

\subsubsection{Measurements at 2.5 K}

\begin{figure}[htb] 
  \centering
  \includegraphics[bb=15 139 602 627,width=3in,height=3in,keepaspectratio]{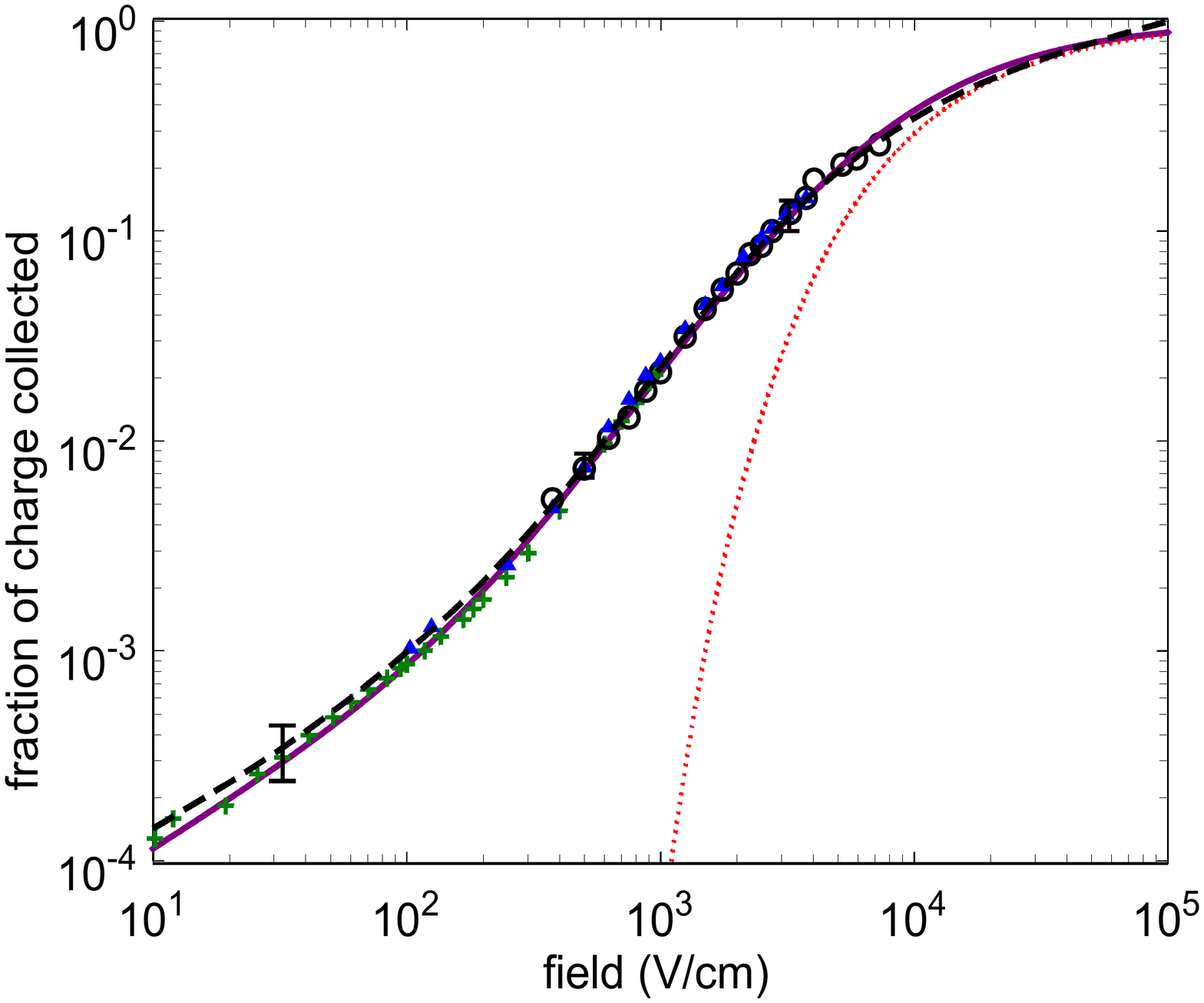}
     \caption{(color online) Current from $^{63}$Ni source as a function of applied field. Symbols are data from measurements in two different cells and at different times with one of them. Representative error bars are shown for several points.
 The dashed and dotted lines are calculations of the escape probability using the Onsager theory of geminate recombination. The dashed line uses the distribution of charge separations shown in Fig.~\ref{fig:dist} while the dotted line is for a Gaussian distribution 
with $ b=4\times 10^{-6}$ cm and no tail. }
  \label{fig:current}
\end{figure}

The results of measurements \cite{Ghosh,Seth} of the current produced by the  beta emitter in liquid helium at 2.5~K are plotted in Fig.~\ref{fig:current} where the current has been normalized to the saturation current (2900 pA) measured in dilute gas. The solid curve is an empirical fit to the data points. The inaccuracies in estimating the derivatives of current can be large  at high fields where no data exists and an extrapolation is required. \\
\vspace{.3cm}

\vspace{.7cm}

\begin{figure}[tb] 
  \centering
  \includegraphics[width=3.5in,height=2.5in,keepaspectratio]{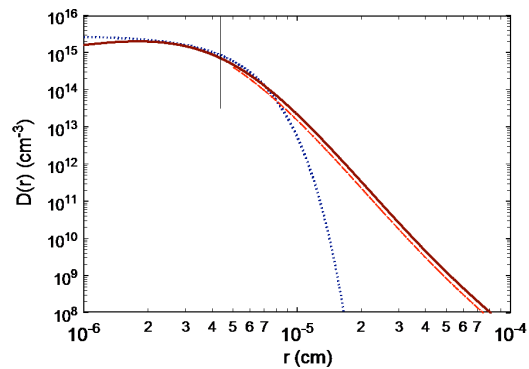}
     \caption{(color online) Solid line: distribution determined from field dependence of current using Eq. (\ref{eq.dis}). Dotted line: Gaussian distribution $\exp(-r/b)^2)$  with $b=4\times 10^{-6}$~cm. Dashed line: Distribution of electrons from track assuming all triplet excimers undergo Penning ionization. The vertical line at $ r= 4.3 \times 10^{-5}$~cm demarks the position given by $r = (e/4\pi\epsilon_0 {\mathcal E})^{1/2}$ corresponding to the high field separating regions where the current has been measured from that where it is extrapolated. The value r of $1.2\times 10^{-4}$~cm marking the low field limit of current measurement is off scale in this plot. }
  \label{fig:dist}
\end{figure}

The charge distribution computed from the empirical fit using Eq. (\ref{eq.dis}) is shown in Fig.~\ref{fig:dist}. Also shown in the figure is a Gaussian distribution, $ D(r)\propto e^{(r/b)^2} $,  
with $b=4\times 10^{-6}$~cm for comparison. Several features are worthy of note. Firstly, the slight decrease in the calculated density at distances below $3\times 10^{-6}$~cm is of no significance. It can easily be the result of a small error in the normalization and/or the extrapolation of the empirical equation into the field region above $10^4$~V/cm where no measurements of current are available. Secondly, the current at low fields bears little relation to what is expected for a Gaussian distribution. The distribution obtained from the data does not decrease exponentially with distance but rather varies approximately as a power law, at least over a limited range in $r$. At large distances the distribution in Fig.~\ref{fig:dist} has approximately an $r^{-6}$ dependence. The distribution has, in the jargon of probability theory, a "fat" or "long" tail.\\

This dependence cannot be attributed to diffusion, which has not been included in the analysis. This can be verified in a number of ways. 1) The Onsager radius $R = e^2/(\epsilon k_B T)$, the separation distance between two charges where the Coulomb energy is comparable to the thermal energy, is larger than $6\times 10^{-4}$ cm at liquid helium temperatures. The escape probability of charges separated by more than this distance are strongly dependent on diffusion whereas those that have separation less than $R$ are not, the latter condition being the case in this analysis as seen in 
Fig.~\ref{fig:dist}. 2) A Monte Carlo calculation of the influence of diffusion on the field dependence of the current assuming a Gaussian spatial charge distribution has been found previously\cite{Guo,Ito} 
to be inadequate to explain the current at low field. 3) A direct measure of the magnitude of the contribution of diffusion to the escape probability can be obtained by inverting the analysis, namely, using the calculated distribution shown in Fig.~\ref{fig:dist} to compute the current dependence on field employing the Onsager solution. The result of performing this calculation, using the Que and Rowlands\cite{Que} expansion of the Onsager solution, is shown as the dashed line in Fig.~\ref{fig:current}. The difference between the solid and dashed lines in the figure indicates the possible error in neglecting diffusion in our analysis. The effect of diffusion only appears at low currents, that is, for escapes which result from charge pairs with large initial separations. The current dependence on field for a Gaussian distribution with $b=4\times 10^{-6}$~cm and no tail, also computed using the Que and Rowlands\cite{Que} expansion, is plotted as the dotted curve in  Fig.~\ref{fig:current}. The striking difference between this curve and the measured current clearly demonstrates that the analysis of the existence of the fat tail is robust, independent of the empirical fitting of the current distribution or extrapolations thereof.  \\

Distributions with long-tail behavior are observed to occur in many physical problems in which rare, outlying events appear more frequently than expected on the basis of a simple model of a random process. In the present case the number of the charges separated by distances larger than expected compared to that for a Gaussian distribution, which fits reasonably well the density profile at small separations, is substantial. The number of electrons with initial separations greater than $10^{-5}$~cm is 10\% whereas the corresponding number for the Gaussian distribution shown in Fig.~\ref{fig:dist} is only 0.5\%.  \\

The characterization of the process by which the electrons undergo thermalization as a random walk from elastic scattering by helium atoms leads naturally to the presumption that resultant spatial distribution is Gaussian. The canonical example of Brownian motion comes to mind. Also,  a random walk on a square lattice in three dimensions results in a Gaussian distribution. However, as is well known\cite{Schle}, in situations where the size of the steps in a random walk is not fixed, and a substantial difference in step length can occur, such as an occasional long step (sometimes referred to as Levy motion), the distribution is non-Gaussian and can have long tails. The random motion of an electron as it thermalizes in liquid helium possesses this characteristic. The distance an electron travels between scatters in liquid helium is highly variable. The single differential elastic scattering cross section, $d\sigma/d \Omega$ depends on energy of the electron and is anisotropic, peaked in the forward direction at energies above 18~eV\cite{Brunger}. Furthermore, the time between steps is expected to be highly variable\cite{Hernandez,Borgh}. The general picture of the mechanism by which an electron becomes thermalized involves the electron in its random walk encountering a region in the liquid where the density is low because of fluctuations. The exchange energy of the non-localized electron with the bound atomic electrons is less in this low density environment. It has been suggested\cite{Hernandez} that the electron could have resonant states in the continuum because of the attractive potential well. The enhanced probability of the electron in a resonant state scattering off atoms would have the effect of deepening the potential well. At some point in an electron's meanderings such a deepened well in a low density region results in a full blown bubble being created. There is no reason to expect that the distribution of electrons from their geminate partners should be Gaussian.\\  

That an energetic electron in losing energy by undergoing elastic scattering in helium does not conform to the standard picture of a random walk has been demonstrated in numerous experiments. For, example, Onn and Silver\cite{Onn69,Onn71}, in studying the behavior of hot electrons injected into liquid helium from a tunnel diode, observed that thermalization occurred over distances and times less than expected from elastic scattering. The current determination of the charge distribution of secondary electrons in liquid helium is but another measurement requiring for its understanding an analysis of how an electron becomes localized within a bubble.\\

There are several sources of free charges that are not directly related to the separation of the initial ionization by the electric field but result secondarily from the stopping of an ionizing particle in liquid helium. The number of these additional charges is not sufficient to affect appreciably the analysis of the initial charge distribution, but we discuss them as a possible source of minor error. \\
\noindent 1) {\it Photoemission}. When an excited helium atom or excimer, He$_2^*$ \cite{comment}, radiatively decays to the ground state, it does so with the emission of a photon with a energy between 13 and 20~eV. Upon hitting a metal surface in vacuum such a photon has a probability in the range of 5 to 15\% \cite{Walker,Hirata} of ejecting an electron, depending on the material and surface conditions. In liquid helium the situation is much different.  From measurements, Guo {\it et al.}\cite{Guo09} estimated the probability of a 16~eV photon creating an electron that escapes from a metal surface into the liquid to be the order of $10^{-4}$. Electrons of several eV are stopped close to the surface and are returned to it because of the image charge, unless there is a sufficient field, as at the cathode of the cell, to produce separation. Given the geometry of our cells, we  expect that at low fields a current the order of 0.2~pA, or normalized current of $5\times 10^{-5}$, to be generated by photoemission. Furthermore, if such a current any larger than this were present, then there should be a noticeable change in current on reversing the polarity of the electrodes in the cells, something that is not observed over that due to the primary electrons from the $^{63}$Ni source.\\
\noindent 2) {\it Auger ejection by He$^+$ and He$_2^+$}. A positive ion with a high ionization potential can, on interacting with a surface, become neutralized with the ejection of an electron by the Auger process\cite{Hagstrum}. Much less is known about the interaction of the ionized dimer He$_2^+$ with a surface but assumed it behaves in a similar way as the single ion. The probability of electron ejection in vacuum can be large, ranging from a few percent to 30\% depending on the surface. But again, in liquid helium an electron will not diffuse far and will return to the surface under the influence of its image charge. Only in the field next to the cathode will the electron have a chance to escape. The contribution to the current from such electrons is expected to be no greater than those from photoemission, and, to the extent that they are produced, they contribute only at high fields where recombination is small. \\
\noindent 3) {\it Excimers}. The He$^+$ ion or an excited atom He$^*$ quickly ($\sim 10^{-11}$~s) form a dimer with a ground state helium, the binding energy being close to 2~eV. While spin-singlet excimers decay to the dissociated ground state in the order of $10^{-8}$~s, the triplet excimer in its lowest energy state, He$^*_2$(a$^3\Sigma_u$), has a radiative lifetime of 13~s\cite{McKinsey99}. When two of these excimers encounter one another in diffusing through the liquid, they interact, producing an ion/electron pair through the exothermic Penning process 
\begin{equation}
{\rm He}_2^* + {\rm He}_2^*   \rightarrow  2{\rm He} + {\rm He}_2^+ + e^-\ .
\end{equation}
The electron has an energy of roughly 10~eV and must thermalize in the same manner as those produced in the initial ionization event. Hence, the distribution of electrons from their geminate positive ions generated by the Penning annihilation of excimers  is the same as the distribution from that of the primary ionization. The fact that the number of excimers depends upon recombination and therefore on applied electric field does lead to a correction in the relationship between field dependence of the current and the density distribution as expressed by Eq.(\ref{eq.dis}), however, the effect is small even at low electric fields where the contribution from Penning ionization is largest.
 The expected ratio of the number of triplet excimers  at low field to the total number of ionizations is 50\%,\cite{Adams}.
 A numerical calculation assuming the maximum possible influence of excimers, that all metastable excimers undergo Penning ionization rather than radiatively decay, results in the size of fat tail of distribution being decreased somewhat but its existence and shape are unaltered, see Fig.~\ref{fig:dist}.
Excimers upon hitting the wall of the cell can also produce electron via the Auger mechanism, but at 2.5~K, where the mobility is low, this is of no consequence. 
\vspace{.2cm}

\subsubsection{Temperature dependence}

\begin{figure}[tb] 
  \centering
  \includegraphics[bb=8 179 606 611,width=3in,height=2.17in,keepaspectratio]{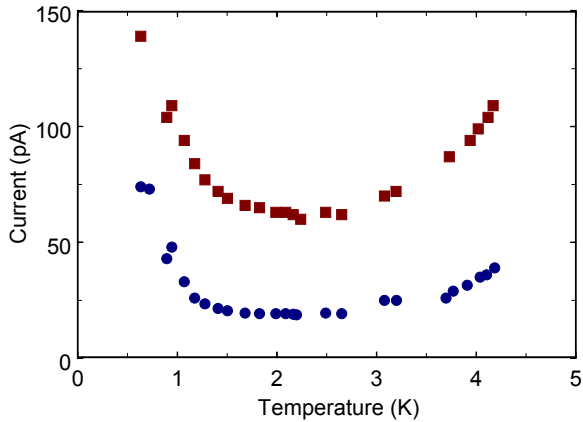}
  \caption{(color online) The temperature dependence of the current at two different fields. Squares: 1000~V/cm. Circles: 500~V/cm.}
  \label{fig:IvsT}
\end{figure}

Measurements were made of the field dependence of the current at other temperatures but only at fields below 1000~V/cm. At low fields the current is observed to decrease by about 40\% with decreasing temperature between 4~K and 2~K, independent of field, see Fig.~\ref{fig:IvsT}. While the magnitude of the current varies with temperature, it maintains the same functional dependence on the field. The variation is much larger than the change in density of the liquid in this temperature region ($\sim$15\%), which affects the range of electrons by their elastic scattering. This suggests that the process of the electron becoming localized by digging a hole in the helium is dependent on density.\\

\begin{figure}[tb] 
  \centering
  \includegraphics[bb=9 153 606 639,width=3in,height=2.44in,keepaspectratio]{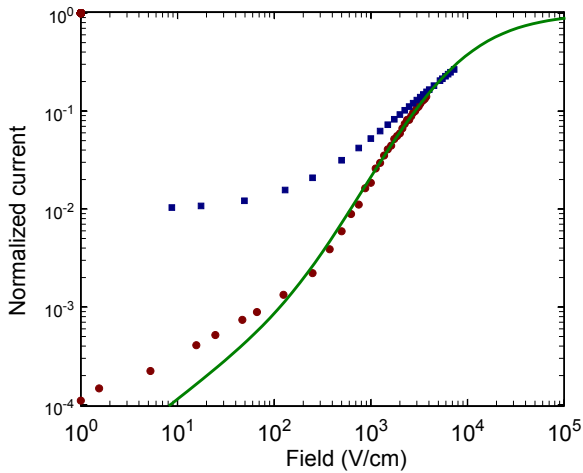}
  \caption{(color online) Line: Empirical fit to the current at 2.5~K. Squares: current at 0.1~K in natural helium. Circles: current at 0.1~K in helium with 6\% $^3$He concentration.}
  \label{fig:compare}
\end{figure}

In the superfluid state of helium an electron or ion, if the field is sufficiently high, creates a vortex ring to which it remains attached\cite{Borgh}. The field at which vortex creation occurs depends on temperature because of the drag on the charge from normal excitations in the liquid. The velocity of a charge attached to a vortex ring has an unusual field dependence, dropping sharply with the creation of the ring and then remaining at a low but slightly increasing value at higher fields. \\

Below 1.2~K the current increases at fixed field with decreasing temperature down to about 0.6~K, see Fig.~\ref{fig:IvsT}. It then remains constant below 0.6~K \cite{Seth}. The field dependence of the current at 0.1~K is plotted in Fig.~\ref{fig:compare} for natural helium and for helium with a 6\% concentration of $^3$He. The current at 0.1~K and 6\% concentration is, to the accuracy of the measurements, indistinguishable from that at 2.5~K above 100~V/cm. On the other hand, for natural helium there is a significant increase in current at low fields, an increase which diminishes as the field increases. The origin of this increase is uncertain. It could be related to the change in motion of the charges because of vortex creation. 
Possibly, it is a consequence of the increased mobility of the long-lived, metastable, triplet He$^*_2$(a$^3\Sigma_u$) excimers at low temperatures. Or it may be the effect of the low viscosity and the resulting increase in escape probability due to the inertia of the bubbles and snowballs\cite{Guo}. Eq.~(3) is derived assuming the velocity is proportional to the force on the charges, which is not the case when the mobility of the charges is high and the acceleration term becomes important. When viscous drag is absent, the condition for escape is not given by Eq.~(3) but rather by $r_0^2 \geq e/(4\pi\epsilon_0\ \mathcal E) .$ \\ 


\vspace{.2cm}

\subsection{Helium gas}\vspace{.2cm}

When the density of helium gas is sufficiently high and the electric field is sufficiently weak, an electron forms a bubble in the gas phase\cite{Sanders}. In thermal equilibrium an electron undergoes as a function of gas density a transition from delocalized state to localized bubble state in the range of density from 1 to $1.5\times 10^{21}$ atoms/cm$^3$ \cite{Sanders} for temperatures below 4.2~K. Also, at gas densities where an electron is in a localized state at zero field it becomes delocalized in high fields\cite{Schwarz}. At a density of $2\times 10^{21}$ cm$^{-3}$,  the transition from localized to delocalized state occurs in the field region above 100~V/cm. Because the mobility is not a constant independent of velocity in the high-field region, Eq.~(\ref{eq.dis}) is not applicable to the field dependence of the current in the gas above 100~V/cm. \\

Neither is Eq.~(\ref{eq.dis}) a particularly good approximation below 100 V/cm. Because the density in the gas is an order of magnitude less than that in the liquid, the separation of geminate partners is on average larger by the same factor, and diffusion plays a more important role. All one can safely conclude from the measured current versus field curves is that the distribution at large separations cannot be described as Gaussian.\\

At a gas density of $0.5\times 10^{21}$cm$^{-3}$ the field dependence of the current differs substantially from that at densities where bubble formation occurs. This observation is consistent with electrons being in delocalized states at low densities.\\
\vspace{.2cm}

\subsection{Alpha particles}\vspace{.2cm}

\begin{figure}[tb] 
  \centering
  \includegraphics[bb=8 159 605 631,width=3in,height=2.37in,keepaspectratio]{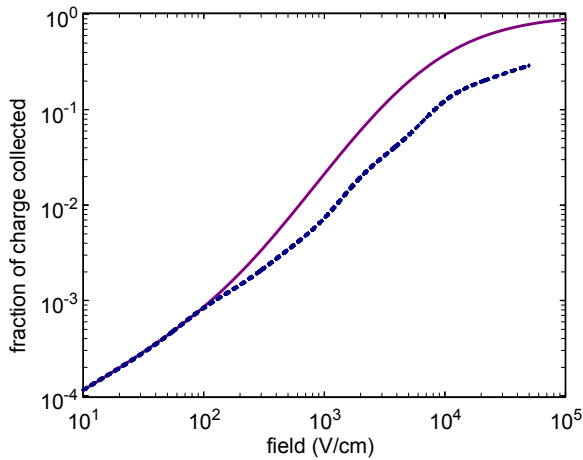}
   \caption{(color online) Solid line: the analytic fit to the measurements of current versus field for betas as illustrated in Fig.~(\ref{fig:current}.  Dashed line: measurements of Williams and Stacey\cite{Williams} of the current produced by alpha particles stopped in liquid helium. The current for alphas is normalized such that the fraction of current is the same for betas at 10~V/cm.}
  \label{fig:alpha}
\end{figure}

The current generated by $\alpha$ particles stopped in liquid helium has been measured\cite{Gerritsen,Williams} some time ago. While the recombination is certainly not geminate along an $\alpha$ track\cite{Ito2}, the current at low fields, below 100~V/cm, appears to have the same field dependence as that observed for betas as illustrated in Fig.~\ref{fig:alpha}. At higher fields this is certainly not the case.\\ 

There is a possible explanation as to why $\alpha$ particles have the same current versus field dependence as betas at low fields. At fields less than 100~V/cm less than 0.1\% of the charges do not recombine. The recombination time for a pair of charges is strongly dependent on their initial separation. The closer the charges, the less the recombination time. By the time there are only 0.1\% of the charges remaining, pairs must be well separated from other charges and the recombination appears to be between isolated pairs, but not geminate pairs. The positive ions are for the most part initially distributed close to the track of the primary ionizing particle while the electrons that do not recombine at low fields are those furthest away from the track. These 0.1\% of unrecombined charges will have moved somewhat during the time that the closer charges recombine, but their distribution may not be appreciably affected. The field dependence of the current in the region below 100~V/cm is that which corresponds to the fat tail in the distribution.\\

A possible mechanism contributing to the fat tail for an energetic $\alpha$ is the substantial number of secondary electrons produced with energies above the ionization threshold of 24.6~eV. These electrons have energies extending up to several hundred eV. Ito and Seidel\cite{Ito}, based on the work of Rudd {\it et al.}\cite{Rudd92} estimated that for an $\alpha$ close to 50\% of the stopping power in helium above 1~MeV is associated with the production of further ionization by secondary electrons. These energetic electrons have a probability, although small, of creating additional ionizations outside of the dense cloud of ions and electrons about the $\alpha$ track. The recombination of ionization events outside the dense cloud, separated in space from other charges, will likely be geminate and hence similar to those occurring when the primary particle is a beta. The number of such events is small because of cross sections for excitation, ionization \cite{Ralchenko}, and elastic scattering\cite{Adib} of electrons with energies between 25 and 1000~eV are such that they will undergo many random scatters if they are to escape the dense cloud about the track.\\
\vspace{.2cm}

\section{CONCLUSIONS}\vspace{.2cm}

The analysis employed here to understanding the charge distribution and recombination along the track of an ionizing particle is substantially different from the approaches that follow from the work of Jaff\'e and Onsager. The applicability of the analytical method we have developed depends on three conditions. 1) Electrons and positive ions interact only in pairs of opposite charge, and 2) the motion of the charges under the influence of electric fields is governed by a mobility that is independent of the velocity of the charge. 3) Diffusion can be neglected. The first condition is satisfied for the ionization products of a beta stopped in liquid helium where the recombination is geminate. The second condition is satisfied in liquid helium since the electron forms a bubble and the positive ion a snowball, both of whose mobilities are governed by scattering of quasiparticle excitations. And the third condition is a result of the temperature being low.\\

The charge distribution about an electron track is found to be approximately a Gaussian distribution having a width of $4\times 10^{-6}$~cm augmented by a ``fat" tail at large distances. The Gaussian distribution is somewhat narrower than would be expected based on the  elastic scattering of electrons from helium atoms. This is presumably related to the process by which an electron becomes localized in a bubble. The temperature dependence of the current suggests that bubble formation depends on density, but this observation needs further measurement and study. The fat tail to the distribution is not unexpected, in that the nature of electron scattering from helium is one with random step length with occasional long steps.\\
 
The applicability of the present analytic method to ionization in other media would appear to be limited for two reasons. Firstly, the mobility must be independent of field.  In general, quasi free electrons moving in a liquid or gas have velocity-dependent mobilities. An analysis of the current in helium gas is a case in point. At low fields where an electron is localized, the field dependence of the current is consistent with a fat tail distribution at large distances. But if the analysis were extended to high fields, the calculated distribution is found to become negative at small distances, an obviously non physical result. And secondly, diffusion plays an important role in the motion and recombination of charges at higher temperatures but is not included in the present study. \\ 

Notwithstanding the limited applicability of the present analysis, the measurement of the secondary electron distribution produced by an energetic ionizing particle in liquid helium may provide insight into the thermalization distances of electrons in other liquids. As discussed by Freeman\cite{Freeman}, in many high-temperature, low-mobility liquids the electron distribution can be fit empirically by a Gaussian with a power law tail. The similarity to the situation in helium would seem to be more than coincidental. It would be of interest to see if the present analytic approach can be adapted to include diffusion so as to handle the full Onsager geminate recombination model. \\
\vspace{.2cm}

\begin{acknowledgments}\vspace{.2cm}

We appreciate helpful conversations with Y.H. Huang, B. Marston, H. Maris, and W. Guo.  This work was supported by the US Department of Energy and the National Science Foundation.

\end{acknowledgments}\vspace{.2cm}

 \appendix*{\centerline {\bf Appendix}} \vspace{.2cm}

Assume an applied electric field of magnitude $\mathcal E$ makes an angle $\theta$ with respect to a vector connecting a pair of charges of opposite sign,  separated by the distance $r$. Take one charge to be fixed. The force on the other in the radial direction on the other charge is
\setcounter{equation}{0}
\begin{equation}
F_r = e{\mathcal E} \cos \theta - \frac{e^2}{4\pi\epsilon_0\ r^2}\  ,
\end{equation}
and in the perpendicular direction  (in the plane of the applied field and radius vector)
\begin{equation}
 F_{\theta} = -e{\mathcal E} \sin \theta\ .
\end{equation}
In the viscous regime where the velocity of the charge is proportional to the force,
\begin{equation}
 \lambda \frac{dr}{dt} = e{\mathcal E} \cos \theta - \frac{e^2}{4\pi\epsilon_0\ r^2}\  ;
\end{equation}
\begin{equation}
  \lambda r \frac{d\theta}{dt} = -e{\mathcal E} \sin \theta \  .
\end{equation}
When one equation is divided by the other, the result is
\begin{equation}
 \frac{dr}{d\theta} = \frac{e}{4\pi\epsilon_0\ {\mathcal E} r \sin \theta} - r \frac{\cos\theta}{\sin\theta}\ .
\end{equation}
With a change in variables $u =r^2$ and $x=\sin\theta$, and after some manipulation
\begin{equation}
\frac{1}{2}\frac{du}{dx} = \frac{e}{4\pi\epsilon_0\ \mathcal E}\frac{1}{x(1-x^2)^{1/2}} -\frac{u}{x}\ ,
\end{equation}
which can be converted to
\begin{equation}
 \frac{1}{2}\frac{d(ux^2)}{dx} = \frac{e}{4\pi\epsilon_0\ \mathcal E}\frac{x}{(1-x^2)^{1/2}}\ .
\end{equation}
Upon integration  
\begin{equation}
 ux^2 = -\frac{2e}{4\pi\epsilon_0\ \mathcal E}(1-x^2)^{1/2} + C\ ,
\end{equation}
or expressed as a relation between $r$ and $\theta$, both of which are  functions of time, 
\begin{equation}
r^2\sin^2\theta = -\frac{2e}{4\pi\epsilon_0\ \mathcal E}\cos\theta +C \ .
\end{equation}
Take the initial conditions to be $r=r_0$ and $\theta = \theta_0$. Then 
\begin{equation}
C = r_0^2\sin^2\theta_0 +\frac{2e}{4\pi\epsilon_0\ \mathcal E}\cos\theta_0  \ .
\end{equation}
If the two charges do not recombine, then as they separate $r$  becomes large as $\theta $ goes to zero. For this to occur the initial positions and field must satisfy
\begin{equation}
 0 <-\frac{2e}{4\pi\epsilon_0\ \mathcal E} + r_0^2\sin^2\theta_0 +\frac{2e}{4\pi\epsilon_0\ \mathcal E}\cos\theta_0  \ ,
\end{equation}
or 
\begin{equation}
 {\mathcal E} > \frac{2e}{4\pi\epsilon_0\ r_0^2} \frac{(1-\cos\theta_0)}{\sin^2\theta_0}= \frac{e}{4\pi\epsilon_0\ r_0^2}\frac{2}{1 +\cos\theta_0}\ . 
\end{equation}
 

or
 \begin{equation}
r_0^2 (1+\cos\theta_0)\geq \frac{2e}{4\pi\epsilon_0\ \mathcal E}.
\end{equation}

\end{document}